\documentclass[preprint]{aastex}

\usepackage{psfig}
\newcommand{\citepeg}[1]{\citep[{e.g.,}][]{#1}}
\newcommand{\citepcf}[1]{\citep[{cf.}\phantom{}][]{#1}}

\def\etal{{\sl et al.}}
\def\lsim{\hbox{ \rlap{\raise 0.425ex\hbox{$<$}}\lower 0.65ex\hbox{$\sim$} }}
\def\gsim{\hbox{ \rlap{\raise 0.425ex\hbox{$>$}}\lower 0.65ex\hbox{$\sim$} }}

\def\f(h{\hbox{$~\!\!^{\rm h}$}}

\def\ale{\mathrel{\hbox{\rlap{\hbox{\lower4pt\hbox{$\sim$}}}\hbox{$<$}}}}
\def\age{\mathrel{\hbox{\rlap{\hbox{\lower4pt\hbox{$\sim$}}}\hbox{$>$}}}}

\def\esi{\hbox{{ESI}}}

\slugcomment{To be published in the {\bf Astronomical Journal} \\received 19
November 2002; accepted 4 December 2002}

\shortauthors{Bloom \etal~  }
\shorttitle{Redshifts of GRB\,990506 and GRB\,000418 with ESI}

\begin{document}

\title{The redshift determination of GRB\,990506 and GRB\,000418
with the Echellete Spectrograph Imager on Keck}

\author{J. S. Bloom\altaffilmark{1,2,3}, E. Berger\altaffilmark{1},
S. R. Kulkarni\altaffilmark{1}, S. G. Djorgovski\altaffilmark{1},
D. A. Frail\altaffilmark{4}}

\bigskip 

\affil{$^1$ Palomar Observatory 105--24, California Institute of Technology,
            Pasadena, CA 91125, USA}

\affil{$^2$ Harvard Society of Fellows, 78 Mount Auburn Street, Cambridge, MA 02138 USA}

\affil{$^3$ Harvard-Smithsonian Center for Astrophysics, MC 20, 60 Garden Street, Cambridge, MA 02138, USA}

\affil{$^4$ National Radio Astronomy Observatory, Socorro, NM 87801, USA}

\begin{abstract}
Using the {\it Echellete Spectrograph Imager} (\esi) on the Keck~II
10-m telescope we have measured the redshifts of the host galaxies of
gamma-ray bursts GRB\,990506 and GRB\,000418, $z=1.30658 \pm 0.00004$
and $1.1181 \pm 0.0001$, respectively.  Thanks to the excellent
spectral resolution of \esi\ ($\lambda/\Delta\lambda = 13000$) we
resolved the \hbox{[O II]} 3727 doublet in both cases. The measured
redshift of GRB\,990506 is the highest known for a dark burst GRB,
though entirely consistent with the notion that dark and non-dark
bursts have a common progenitor origin. The relative strengths of the
\hbox{[O II]}, \hbox{He I}, \hbox{[Ne III]}, and \hbox{H$\gamma$}
emission lines suggest that the host of GRB\,000418 is a starburst
galaxy, rather than a LINER or Seyfert 2.  Since the host of
GRB\,000418 has been detected at sub-millimeter wavelengths these
spectroscopic observations suggest that the sub-millimeter emission is
due to star-formation (as opposed to AGN) activity.  The \hbox{[O
II]}-derived unobscured star-formation rates are 13 and 55 $M_\odot$
yr$^{-1}$ for the hosts of GRB\,990506 and GRB\,000418,
respectively. In contrast, the star-formation rate of the host of
GRB\,000418 derived from sub-millimeter observations is twenty times
larger.

\end{abstract}


\section{Introduction}

The determination of redshifts continue to play a crucial r\^{o}le in the
understanding of the physics of GRBs.  Without redshifts, we are
unable to translate observed fluences and angular burst-host offsets
into physically meaningful quantities.  That redshift is an integral part
of the physical modeling of the afterglow is best demonstrated by the
immense difficulty in modeling the afterglow of GRB\,980329
\citep{yfh+02}, a GRB that enjoys an abundance of pan-chromatic
afterglow observations but unfortunately lacked a measurement of
redshift.

Emission spectra of GRB hosts themselves are valuable given the
potential use of GRBs to infer the star-formation history of the
Universe. Since $\gamma$-rays penetrate dust, a GRB-defined sample
does not suffer from the well known biases of optical/UV samples. In
addition, the high-accuracy astrometric localizations afforded by
radio observations (with, e.g., the Very Large Array; VLA) and/or
X-ray (Chandra) offer significant observational advantages over a
sub-millimeter-defined sample. Indeed, one great hope for the utility
of GRBs is in revealing the star-formation history at extremely high
redshifts \citepeg{bl02} since the brilliance of the prompt emission
allows these events to be seen to redshifts well beyond 10
\citep{lr00}.

GRB\,990506, is a canonical ``dark
burst,''\footnotemark\footnotetext{GRB\,970828 is another such dark
burst, with a similar detection history as GRB\,990506 \citep{dfk+01}:
later optical observations showed that the radio-localized host of
GRB\,970828 was a dusty galaxy at $z\sim 1$.} with a radio-detected
afterglow with no optical transient emission
\citep{tbf+00}.  Since no transient optical emission was detected from
the afterglow, an optical absorption-line redshift of GRB\,990506 was
impossible to obtain.  \citet{tbf+00} discovered an apparent
two-galaxy system in optical data from Keck consistent with the radio
transient position.  Subsequent imaging from the {\it Hubble Space
Telescope} (HST) revealed the putative two-galaxy system to be
morphologically distinct \citep{htab+00}. \citet{bkd02} later refined
the radio$\rightarrow$HST astrometry to show that the radio afterglow
position is only consistent with the southwest component; this
component is identified as the host of GRB\,990506. However, given the
poor astrometric tie between the radio and the optical positions, as
noted by \citeauthor{bkd02}, the GRB\,990506--galaxy association is
one of the more tenuous identifications of a host galaxy.

GRB\,000418 is famous for two reasons. First, it is still one of the
best cases for collimated ejecta of a GRB explosion \citep{bdf+01}.
More important, the host galaxy has been detected in the radio (using
the Very Large Array, VLA) and the sub-millimeter bands (using the
Sub-millimeter Common User Bolometer Array, SCUBA, on the James Clerk
Maxwell Telescope); see \citet{eck+02}.  If the sub-mm emission arises
from reprocessing of stellar UV light by dust (as opposed to arising
from AGN activity) then the inferred star-formation rate is very
high. If so, this galaxy can potentially provide information about the
nature of these ultra-luminous galaxies and their contribution to the
global star formation rate.

Here we present the spectroscopic observations of the hosts of
GRB\,000418 and GRB\,990506. The discussion of GRB\,000418 here
expands our preliminary announcement of the redshift of the host
\citep{bdd+00}.  In \S
\ref{sec:esiobs} we describe the observations and reductions that led
to the redshift determinations (presented in \S\S
\ref{sec:0506redshift}--\ref{sec:0418redshift}).  In \S
\ref{sec:natureofhosts} we discuss some of the implications of the
redshift detections in the context of GRB host galaxies.

\section{Observations and Reductions}
\label{sec:esiobs}

The observations of both GRB host galaxies were obtained in the
echellete mode of the {\it Echellete Spectrograph Imager}
\citep[\esi;][]{sbe+02} on the Keck II 10-m telescope atop Mauna Kea,
Hawaii.  The detector is a $2{\rm k}\times 4{\rm k}$ pixel Lincoln
Labs CCD with low readnoise (r.m.s.~$\approx$ 2.6 electron) and small
dark current (2.1 electron pixel$^{-1}$ hr$^{-1}$). The wavelength
coverage is complete from $\sim 3900$ \AA\ to 1 $\mu$m spanning 10
echellete orders (orders 15 through 6, increasing with wavelength).
The (inverse) dispersion is nearly constant at 11.5 km s$^{-1}$
pixel$^{-1}$.  For all of our observations we used the 1.0 arcsec
$\times$ 20.0 arcsec slit.

The putative host galaxies were identified as the galaxies nearest the
astrometric position of the GRB afterglows.  \citet{bkd02} estimated
the chance that the afterglows were unrelated to the presumed host at
$P_{\rm ch} = 0.044$ (GRB\,990506) and $P_{\rm ch} = 0.0004$
(GRB\,000418). The brightness of these galaxies are $R = 24.0 \pm 0.3$
mag \citep[GRB\,000418;][]{bdg+00} and $R = 24.3 \pm 0.3$ mag
\citep[GRB\,990506;][]{tbf+00}. Both the galaxies are compact
(half-light radius $<$ 0.2 arcsec) as measured by HST/STIS
observations \citep{mfm+00,htab+00,bkd02}.  A summary of the
observations of both GRB hosts is presented in Table
\ref{tab:esilog}. For both sources, we centered the slit 
on a bright nearby star and then offset the telescope to the host
galaxy by a distance derived from our ground-based astrometric
solution of the field.  Exposures of the listed spectrophotometric
standard stars from
\citet{oke90} were obtained immediately following the last GRB
host exposure. All science and calibration data were taken near the
parallactic angle.

The data were reduced and calibrated using the standard spectral
reduction packages in IRAF\footnotemark\footnotetext{IRAF is
distributed by the National Optical Astronomy Observatories, which are
operated by the Association of Universities for Research in Astronomy,
Inc., under cooperative agreement with the National Science
Foundation.}, primarily the packages CCDRED, NOAO/ONEDSPEC and
ECHELLE.  For completeness we outline the steps taken.  The data were
first debiased using the overscan region of the CCD and then bad
pixels were replaced using FIXPIX.  A flat field was constructed from
five domeflats taken at the beginning of the night and applied to the
science data.  Cosmic-rays were identified by finding statistically
significant high-valued pixels in the ratio of each science image with
the median of the science images.  The cosmic ray pixels were then
replaced using FIXPIX. Using IDENTIFY, REIDENTIFY, and FITCOORDS we
characterized the curvature of the 10 echellete orders using a trace
of a bright star which we stepped along the slit with a constant two
arcsecond offset. A plate scale in the spatial direction for each
echellete order was also obtained using these spectra.  Rectification
(setting the dispersion and spatial axes to be orthogonal) of the
science image was performed using the task TRANSFORM with flux
conservation.

In each image, the 10 orders were then cut into individual longslit
2-d spectra.  The background sky was removed from each spectra using
BACKGROUND.  Since a feature near 8600 \AA\ was readily identifiable
(see below) in order seven (8006 \AA\ $< \lambda < $ 9377 \AA) of each
exposure of GRB\,990506 we coadded the background subtracted spectra
in order seven using IMCOMBINE by determining the offset in pixels
along the spatial direction of this feature from exposure to exposure.
Spectrograms in the other orders were combined by scaling the offsets
determined in order seven using the empirically-determined spatial
scales in each order. Similarly, for GRB\,000418, we determined
offsets using the emission feature at 7900 \AA.

The continuum of the host of GRB\,990506 is not detected in the blue
(orders 15 to 9 corresponding to 3756 \AA\ $< \lambda < $7297 \AA) and
marginally detected in the three remaining red orders. Wavelength
calibration was performed in the red orders using the night sky lines
only.  The density of these lines in the red is sufficiently high that
arc line spectra were not required for wavelength calibration. In the
seventh echellete order, a good wavelength solution was obtained with a
seven order Legendre polynomial which resulted in an
r.m.s.~uncertainty of 0.16 \AA. The small air-to-vacuum and
heliocentric corrections were made using DISPTRANS, RVCORRECT, and
DOPCOR.  The flux calibration using the standard star observations
were performed using STANDARD, SENSFUNC, and CALIBRATE.

We produced a final co-added flux calibrated spectrum of the host of
GRB\,000418 using the five 1200-s exposures (Table 1) and following
the reduction procedure as described above for GRB\,990506. We
dereddened the 000418 host spectrum using the extinction measure of
$E(B-V)$ = 0.3296 from \citet{sfd98}.  Unlike the case for
GRB\,990506, since the night was not photometric, a zero-point
adjustment was made using the calibrated and extinction corrected Gunn
i magnitude (i = 23.38 $\pm$ 0.10 mag) from
\citet{bdf+01} at the time of the spectroscopic observations.
Using the zeropoint from \citet{fsi95}, the continuum flux at the
effective wavelength of the Gunn i (for a spectrum with $f_\nu
\propto \nu^{-2}$) should have been $f_\nu(7973$\,\AA) $= 2.12\,\mu$Jy
whereas the measured value is 1.10 $\mu$Jy.  We thus scaled the
spectrum by a factor of 1.93 and estimate a zeropoint error of 30\%.

\section{The Redshift of GRB\,990506}
\label{sec:0506redshift}

We visually identified an apparent doublet emission feature near 8600
\AA\ in order seven of each 900s spectrum.  The spatial location of
this feature was used to coadd the dithered spectra as described
above.  Unfortunately this apparent doublet is separated by a sky line
(O-H 8597.4\,\AA). However, given that the sky line and the fringing
at this wavelength are both relatively weak, we are confident that the
sky subtraction did not significantly alter the morphology of the
feature; that is, we believe the feature is indeed a doublet (such as
[O II]) and not a single broad emission line (such as Lyman
$\alpha$). In fact the sky line only overlaps the redward wing of the
bluer emission line; the shape and flux in the red emission line is
largely unaffected by the sky line.  A portion of the 2-dimensional
spectrogram of the host of GRB\,990506 is displayed in Figure
\ref{fig:spect0506}.  The \hbox{[O II]} emission doublet is 
clearly visible as is the faint continuum.

Using the $\chi^2$-minimization deblending algorithm in IRAF/SPLOT we
preliminarily fit two Gaussian profiles plus a continuum to the
emission features in surrounding region allowing all the parameters to
be fit independently. The central wavelengths of the two emission
lines was found to be $\lambda_1$ = 8596.80 \AA\ and $\lambda_2$ =
8603.25 \AA.  If we associate these lines with [O II] $\lambda\lambda$
3727.092, 3729.875 then the implied redshift is $z_1$ = 1.30657 and
$z_2$ = 1.30658.  That these two redshifts agree, greatly strengthens
the association of these lines with [O II] emission.

Given the interpretation of the emission features as due to [O II], we
refine the redshift determination by fixing the relative separation of
the two features to be the ratio of the [O II] doublet {\it in
vacuo}. Since the blueward line was closer in wavelength to the sky
line we also fix the Gaussian FWHM of the blueward feature to be that
of the redward feature (3.9 pixels).  The emission redshift of the
host of GRB\,990506 so determined is $z(990506) = 1.306576 \pm
0.000042$.  This includes the statistical uncertainty in the
wavelength solution and the wavelength centering of the doublet.

A confirmation of the redshift using other emission lines proves
difficult as most of the other strong nebular line emission features
associated with star formation (H$\alpha\ \lambda$6563, H$\beta\
\lambda$4861, [O III] $\lambda$4363, $\lambda$ 4959, $\lambda$ 5007) are
redshifted outside the \esi\ bandpass.  However, the detection of the
galaxy in the R-band \citep{tbf+00} implies that the feature could
certainly not be Lyman $\alpha$ $\lambda$1216.

Assuming a flat CDM cosmology \citepeg{dab+00} with $H_0 = 65$ km
s$^{-1}$ Mpc$^{-1}$, $\Omega_M = 0.3$, and $\Lambda_0 = 0.7$ the
luminosity distance to the galaxy is $3.056 \times 10^{28}$ cm and 1
arcsec is 9.03 kpc (physical) in projection.

\section{The Redshift of GRB\,000418}
\label{sec:0418redshift}

We detect a strong doublet at $\lambda \approx 7898$\,\AA\ and
identify this doublet with the \hbox{[O II]} doublet due to emission
from the underlying host galaxy at redshift $z \simeq 1.1181$. A
subsequent search for associated star-formation emission lines reveals
the presence of three faint lines (He I, [Ne III], H\,$\gamma$) in the
spectrum which corroborate the redshift.  The detection of H\,$\gamma$
is marginal relative to the other detections. The identifications of
the lines are listed in Table \ref{tab:emline0418}. The spectrum in
the vicinity of [O II] and He I \& Ne III are shown in Figure
\ref{fig:spect0418}.  Unlike in the spectrum of GRB\,990506, the [O
II] doublet is not particularly affected/overshadowed by a neighboring
sky line, although the proximity of a faint sky line just blueward of
the 3727.11\,\AA\ line adds to the uncertainty in the measured
equivalent width (EW) and luminosity.

The luminosity weighted redshift using the [O II], He I and [Ne III]
lines is $z(000418) = 1.1181 \pm 0.0001$.  Using the cosmology assumed
above, the luminosity distance to the galaxy is $2.520 \times 10^{28}$
cm and 1 arcsec is 8.83 kpc (physical) in projection.

\section{The Nature of the hosts galaxies}
\label{sec:natureofhosts}

\subsection{The Host Galaxy of GRB\,990506}

We estimate the unobscured star formation in the host by measuring the
flux of the \hbox{[O II]} line. First, we assess the accuracy of our
flux calibration.  At 8600 \AA\ we find that the flux of the continuum
is 0.57 $\mu$Jy with a red continuum, the slope of which is difficult
to fit over a large wavelength range. Taking the continuum spectral
index to be $0.2 \ale \alpha \ale 0.8$, with $f_\nu \propto
\lambda^\alpha$, the flux at the central wavelength of the $R_C$
filter is $0.54 \age f_\nu(6415$ \AA) $\age 0.45 \mu$Jy. Using the
zeropoint from \citet{fsi95}, the corresponding R-band spectroscopic
magnitude is $24.37 \ale R_C ({\rm spect}) \ale 24.57$ mag. This
spectrophotometric flux estimate agrees reasonably well with the
photometric measurement of $R = 24.3 \pm 0.3$ mag from \citet{tbf+00}.
From the difference ($\approx$ 0.1 mag) we estimate a conservative
systematic uncertainty of our flux calibration at 20\%.

Using the line-deblending method described above we found the observed
flux in the emission doublet to be $F_{8596} = (7.01 \pm 1.2) \times
10^{-18}$ erg s$^{-1}$ cm$^{-2}$ and $F_{8603} = (12.47 \pm 1.1)
\times 10^{-18}$ erg s$^{-1}$ cm$^{-2}$.  According to the dust maps
of \citet{sfd98}, the Galactic extinction in the direction of
GRB\,990506 is $E(B-V)$ = 0.0686 mag. Assuming the $R_V = 3.1$
extinction curve of \citet{ccm89} the dereddened flux at 8600 \AA\ is
10.7\% higher than observed.  The total dereddened flux in the [O II]
line is thus, $F_{{\rm [O II]}} = (2.16 \pm 0.18) \times 10^{-17}$ erg
s$^{-1}$ cm$^{-2}$.

The total luminosity in the [O II] line is $L_{\rm [O II]} = (2.53 \pm
0.21) \times 10^{41}$ erg s$^{-1}$. We have no {\it a priori}
determination of the intrinsic extinction due to dust external to the
Galaxy so this luminosity should be considered a lower limit.  The
implied unobscured star formation rate in the GRB host galaxy is,
using the relation from \citet{ken92}, 12.6 $M_{\odot}$ yr$^{-1}$.
This estimate is uncertain by $\sim$10\% (statistical) and 20\%
(systematic).

In the restframe, the total equivalent width is EW$_{{\rm [ O II]}} =
36.5 \pm 5.44$ \AA. No comparison sample of field galaxy [O II] EW
measures at comparable redshifts has yet been published, but it is
clear that the EW is not especially higher than an extrapolation from
the low-redshift sample of \citet{hcbp98} [Aside from the redshift
  differences, we caution against the use of the Hogg sample as a
  representative sample for comparison to GRB hosts; galaxies in that
  sample were selected on the basis of optical magnitudes and, given
  the trend of decreasing EW with increasing apparent galaxy
  brightness, systematically selects against galaxies of high EW.]
Nevertheless, to this end, the star formation rate per unit mass does
not appear to be particularly high for star forming galaxies at
comparable redshifts.

The observed EW in the two [O II] lines are EW$_{3727.11} = 30.6 \pm
7.9$ \AA\ and EW$_{3729.86} = 53.6 \pm 9.7$ \AA. The ratio of the
lines EW$_{3729}$/EW$_{3727} = 1.75 \pm 0.55$. This is consistent with
the ratio expected (1.5) in the limit of low density in the host
galaxy HII regions \citep{ost74}. Note that the formal error on this
ratio, found assuming the statistical errors quoted above on both
lines, is likely an overestimate of the true uncertainty since the
uncertainty in the continuum level---which is essentially the same for
both lines---dominates the statistical error.

\subsection{The Host Galaxy of GRB\,000418}

The fluxes and equivalent widths of the observed lines are given in
Table \ref{tab:emline0418} and were found using the same deblending
method as above.  None of the lines appear to be extended beyond the
instrumental resolution; that is, all of the lines are narrow. The
total (dereddend) flux in the [O II] line is $F_{[{\rm O II}]} =
(1.4 \pm 0.2) \times 10^{-16}$ erg s$^{-1}$ cm$^{-2}$ corresponding to
a luminosity in the [O II] line of $L_{\rm [O II]} = (1.1 \pm 0.16)
\times 10^{42}$ erg s$^{-1}$. The implied unobscured star formation
rate in the GRB host galaxy is, using the relation from
\citet{ken92}, 55 $M_{\odot}$ yr$^{-1}$.  This estimate is
uncertain by $\sim$10\% (statistical) and 30\% (systematic).

The total observed equivalent width of the [O II] line is EW$_{\rm
obs}$ = 111 $\pm$ 6\,\AA\ corresponding to a restframe EW$_{\rm rest}$
= 52.4 $\pm$ 2.8 \AA. This does appear to be higher than an extrapolation
of the Hogg sample and is suggestive of high star formation per unit
mass.  The ratio of the [Ne III] to [O II] flux is of interest both as
a starburst/AGN discriminator \citepeg{rtt97} and because several
other GRB hosts appear to have a large ratio, indicating very hot HII
regions \citepeg{bdk01}. The observed ratio, $\sim 0.03$, is lower
than that inferred from GRB\,970228 and rules out a Seyfert 2 origin of
emission \citep[following Fig.~4 of][]{rtt97}. With the lines
observed, we cannot directly address whether the galaxy is a LINER
(both H$\beta$ and [O III] are redshifted outside the range of
observability with \esi). However, assuming Case B recombination and
little intrinsic extinction, the implied ${\rm H}\beta$ flux is
$F_{{\rm H}\beta} \approx F_{{\rm H}\gamma}/0.474 = 8.4 \times
10^{-18}$ erg s$^{-1}$ cm$^{-2}$. The implied flux ratios of [O
II]/H$\beta$ and [Ne III]/H$\beta$, then, place this source
comfortably near starburst galaxies and sufficiently away from the
locus of known LINERs \citep[Fig.~3a of][]{rtt97}. We thereby identify
the host of GRB\,000418 as a starburst galaxy.

The radio and sub-mm detections of the host galaxy of GRB\,000418 also
indicate a high star formation rate.  We estimated the star formation
rate using the formulation of \citet{yc02}. As detailed in
\citet{eck+02}, the flux density at 350 GHz, $F_\nu(350\,{\rm
GHz})=3.2$ mJy, implies ${\rm SFR}\approx 700$ M$_\odot$ yr$^{-1}$,
while the flux density at 8.46 GHz, $F_\nu(8.46\,{\rm GHz})=51$
$\mu$Jy, implies ${\rm SFR}\approx 330$ M$_\odot$ yr$^{-1}$.  For the
radio-to-SFR conversion we assume $\alpha=0.6$ \citep{fkp+02}. The
discrepancy between the two SFR estimates is expected since in both
cases there is a $\sim 25\%$ uncertainty in the flux density.  Thus,
the SFR implied from the long-wavelength observations is in the range
$\sim 250-900$ M$_\odot$ yr$^{-1}$. 

Clearly, this SFR estimate is about a factor of $10-20$ higher than
the estimate from the [O II] emission.  This is not surprising given
that the [O II] luminosity is a measure of the unobscured SFR, and
correction factors of $5-10$ for high redshift galaxies are not
uncommon \citep{as00}. Interestingly, there was little to no host
galaxy extinction inferred in the GRB afterglow itself \citep{bdf+01}.

\section{Discussion and conclusion}

Traditionally, low-resolution spectroscopy ($R \approx 1000$) has been
used to determine emission-line redshifts of GRB hosts. Indeed, on the
night previous to the [O II] detection of the host of GRB\,990506 in
echellete mode, a one hour low-resolution long-slit prism spectrum was
obtained and no lines were detected. This non-detection motivated our
subsequent use of the spectrograph in echellete (medium resolution)
mode.

Medium- to high-resolution spectrometers have already been used for
the detection of absorption-line redshifts of GRBs
\citep{cdd+00,cdk+00,mhk+02,sfi+02}, but the redshifts of the hosts of
GRB\,000418 and GRB\,990506 are the first emission-line redshifts
found with a medium-resolution ($R=13000$) echellete spectrograph.

This point is worth highlighting from an observational perspective.
Given that the throughput of \esi\, in echellete mode ($\ale 12\%$) is
approximately 40\% smaller than that of the prism long-slit
(``high-throughput'') mode \citepcf{sbe+02}, it is not immediately
obvious why echellete spectroscopy would be able to detect the
redshift of GRB\,990506 whereas long-slit spectroscopy failed to
detected the line.  While both observing modes provide similar
wavelength coverage and slit widths, however, the prism long-slit
mode results in a lower dispersion with increasing wavelengths (from 1
\AA/pixel at 4000 \AA\ to $\sim$10 \AA/pixel at 10,000 \AA).  This
effect causes the numerous sky emission lines to overlap to the extent
that, beyond about 6000 \AA, almost all pixels are dominated by
dispersed sky emission lines.  In the echellete-mode, the sky lines
are always resolved except for the largest slit widths.  Even if sky
subtraction were perfect, the noise from the high sky background
levels in low-resolution mode more than mitigates, in terms of
signal--to--noise, against the gain in throughput.

\begin{figure*}[tbp]
\centerline{\psfig{file=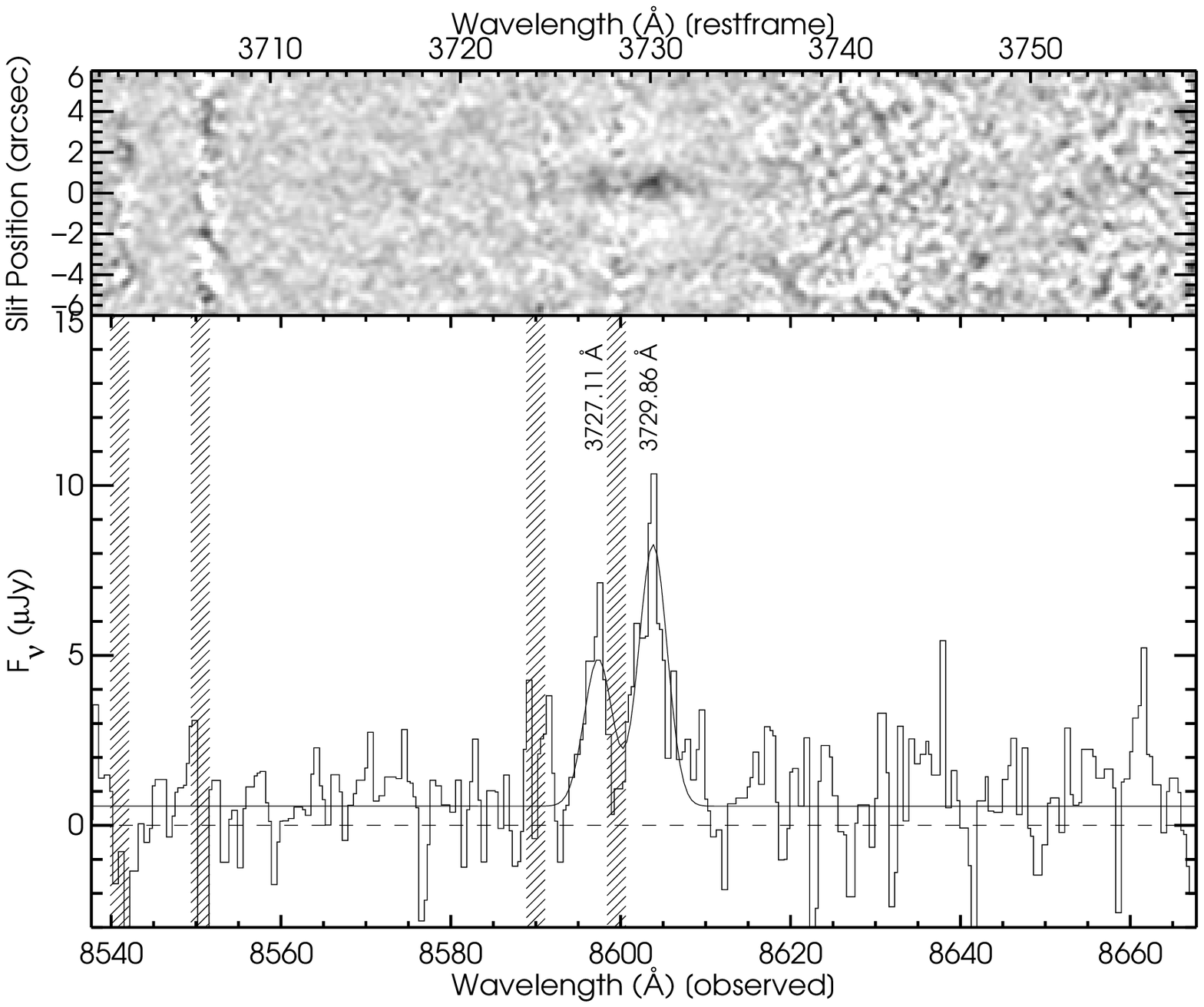,width=7.2in}}
\caption[]{The summed spectrum of the host galaxy of GRB\,990506 in the
vicinity of the [O II] emission doublet.  (top) The background
subtracted two-dimensional spectrogram smoothed by a Gaussian with
$\sigma = 0.5$ \AA.  The doublet is clearly visible at the center of
the spectrogram and the faint continuum is also present. (bottom) The
1-d spectrum. The solid line represents the fit of the continuum plus
[O II] doublet flux to the data with the restframe wavelengths of the
two emission lines noted.  The hashed vertical lines mark the location
of strong night sky lines.}
\label{fig:spect0506}
\end{figure*}

\begin{figure*}[tbp]
\centerline{\psfig{file=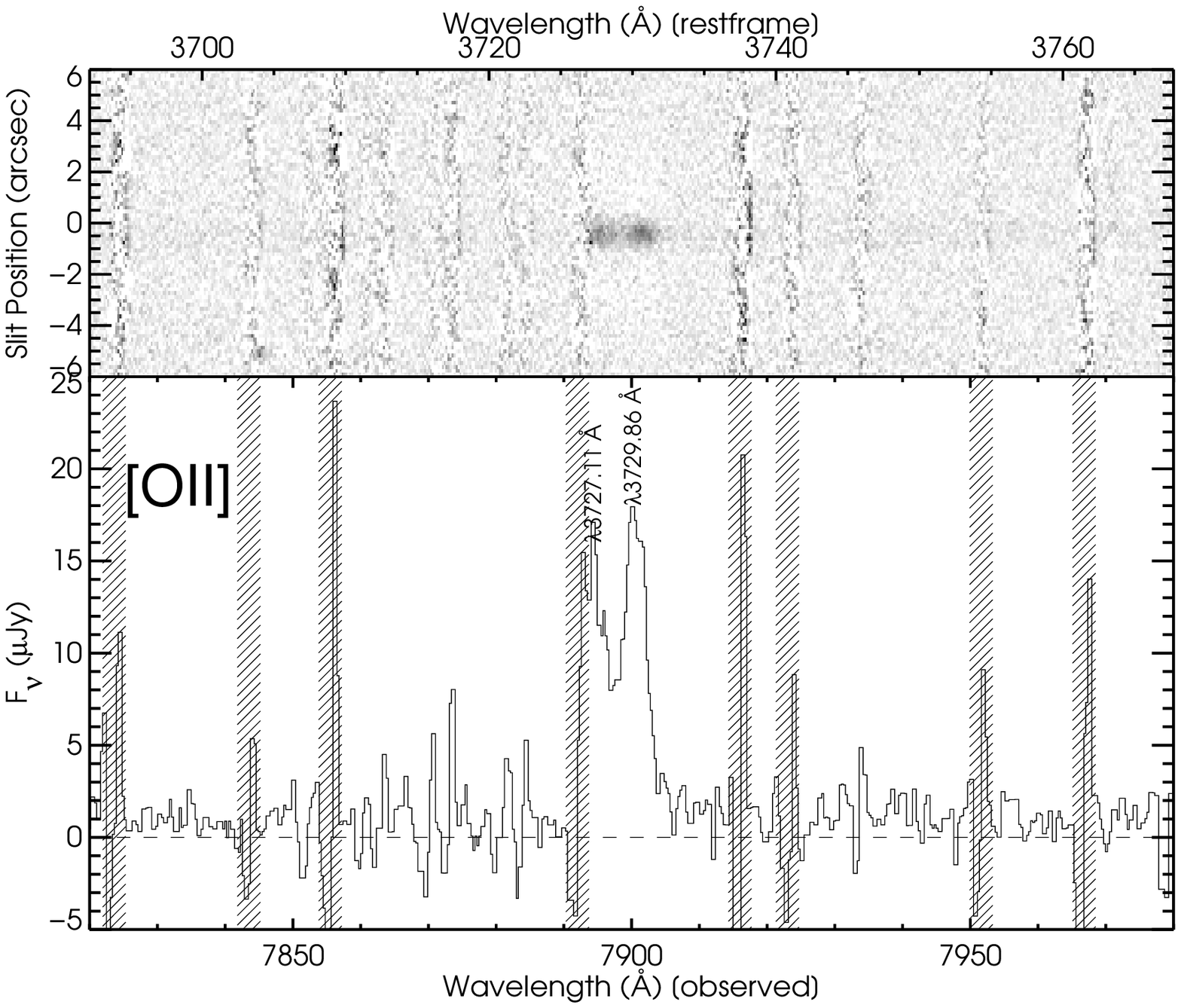,width=4.2in}}
\centerline{\psfig{file=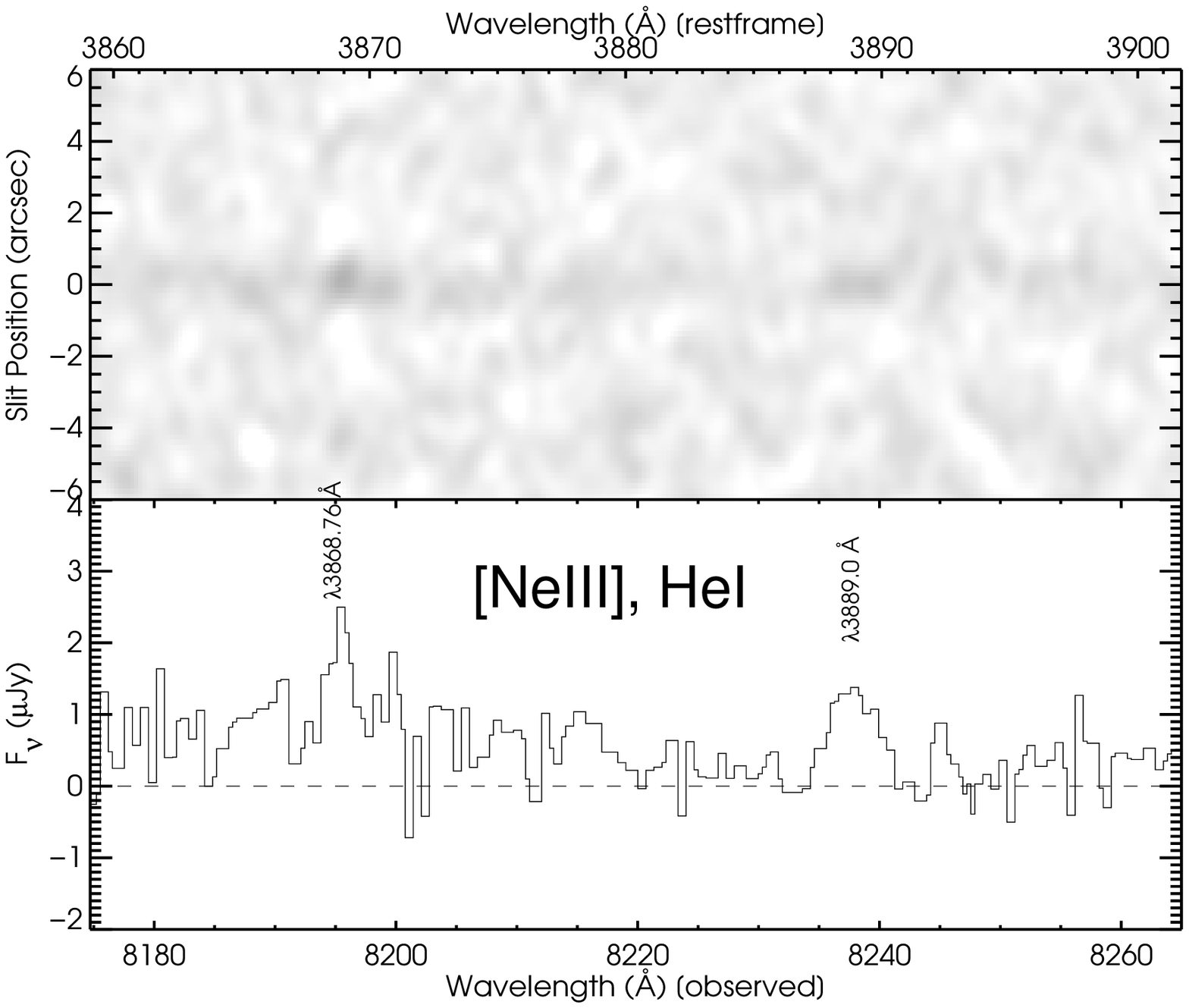,width=4.2in}}

\caption[]{The summed spectrum of the host galaxy of GRB\,000418 in the
vicinity of (top) the [O II] emission doublet and (bottom) [Ne III]
and He I.  The background subtracted two-dimensional spectrogram
smoothed by a Gaussian with $\sigma = 0.5$ \AA.  As in Figure
\ref{fig:spect0506} the doublet is clearly visible at the center of
the spectrogram and the continuum is also detected.  The hashed
vertical lines mark the location of strong night sky lines.}
\label{fig:spect0418}
\end{figure*}

Is it clear from previous studies that GRB hosts contain moderate
levels of unobscured star formation, resulting in strong Balmer line,
[O II] and Ly $\alpha$ emission-lines \citep[see][for review]{dkb+01}.
Most GRBs, however, appear to occur at redshifts near unity and so
H$\alpha$ and Ly$\alpha$ lines are outside the range of detectability
for optical CCDs. At such redshifts, [O II], typically the third most
luminous star-formation line, resides at $\lambda \age 6000$ ($z \age
0.6$) in the observer's frame; however, this is the onset of the
wavelength regime where the number of prominent sky lines becomes numerous
and densely packed.  As a faint, typically narrow line (observed FWHM
$\ale 6$ \AA), the [O II] line can be easily outshone from a nearby
night sky line when the dispersion is low.  This was clearly the case in
GRB\,990506 (Figure~\ref{fig:spect0506}).

Medium-resolution spectroscopy also affords a better insight into the
nature of any emission line detections.  Specifically, the ambiguity
of single line redshifts \citep[see][]{sbs+00} is essentially removed
since the resolution of \esi\, is large enough to resolve the [O II]
doublet (restframe separation of 2.75 \AA) for redshifts greater than
$z \sim 0.5$.

From the optical lines, we have shown that the unobscured
star-formation rates in the hosts of GRB\,990506 and GRB\,000418 are
$13$ and $55$ M$_\odot$ yr$^{-1}$, respectively.  This may be
contrasted with ${\rm SFR}\sim 600$ M$_\odot$ yr$^{-1}$ from sub-mm
and radio observations of GRB\,000418.  Such a discrepancy has been
observed in many high-$z$ galaxies and it is probably due to dust
obscuration. Such a trend, for GRB galaxies, has been previously noted
by comparing infrared photometry with spectroscopic star-formation
indicators \citep{cba02}. More importantly, however, spectroscopy of
GRB\,000418 has allowed us to determine the redshift to a sub-mm
galaxy.  This is a significant result since there are only a handful
of such galaxies with a measured redshift \citepeg{lso+02}.  This
hints at the unique potential of GRB-selected galaxies in uncovering
the redshift distribution of the population of sub-mm galaxies.

Since GRB\,990506 is one of only a handful of well-studied dark bursts
(also, GRB\,000210, \citealt{pfg+02} and GRB\,970828,
\citealt{dfk+01}) the measurement of a redshift near the median of
other long-duration bursts may suggest a common progenitor
population. This emerging trend, of similar redshift distributions,
will almost certainly be testable with the advent of systematic
absorption redshift determinations for {\it Swift} bursts.

\acknowledgments

The authors thank the generous support of the staff of the W.~M.~Keck
Foundation.  JSB gratefully acknowledges the fellowship from the
Fannie and John Hertz Foundation and a research grant from the
Harvard-Smithsonian Center for Astrophysics. SRK's research in GRBs
is supported by NSF and NASA.  SGD acknowledges partial funding from
the Bressler Foundation. We thank S.~Castro for helpful discussions
concerning \esi\ reductions.  The National Radio Astronomy Observatory
is a facility of the National Science Foundation operated under
cooperative agreement by Associated Universities, Inc. The authors
wish to extend special thanks to those of Hawaiian ancestry on whose
sacred mountain we are privileged to be guests. Without their generous
hospitality, the observations presented herein would not have been
possible.

\newpage

\newpage

\newpage

\begin{center}
\begin{table*}
\caption[]{Log of Spectroscopic observations of GRB\,990506 and GRB
000418\label{tab:esilog}}
\begin{tabular}{lcccccr}
\hline
\hline
Name & Date & Int. Time & Airmass & Standard & Photometric? \\
     & (UT) &   (sec)   & (effective) & Star  &             \\
\hline
GRB\,990506 & 4.395 Apr 2000 & 7$\times$900 & 1.52  & Feige 67  & Yes \\

GRB\,000418 & 1.319 May 2000 & 1$\times$1200 & 1.01 & Feige 34  & No  \\
            & 2.403 May 2000 & 5$\times$1200 & 1.2  & Feige 66  & No \\
\hline
\end{tabular}
\raggedright

\end{table*}
\end{center}

\begin{center}
\begin{table*}
\caption[]{Emission line properties of the host of GRB\,000418\label{tab:emline0418}}
\begin{tabular}{lccccr}
\hline
\hline
Line ID & \multicolumn{2}{c}{Wavelength$^a$} & $F(\nu_0)$ & EW$^b$ & $z$ \\
        & Restframe      &  Observed     & [erg cm$^2$ s$^{-1}$] &
\AA\    &  \\
\hline

[O II]   & 3727.11 \AA    & 7894.42 \AA   & 2.8 $\times 10^{-17}$ & $52
\pm 4$  & 1.11811 \\
        & 3729.86 \AA    & 7900.23 \AA   & 3.8 $\times 10^{-17}$ & $69
\pm 5$  & 1.11810 \\

[Ne III] & 3869.85 \AA   & 8196.96 \AA   & 2 $\times 10^{-18}$ &
$\sim 7$ & 1.11816 \\

HeI      & 3888.65 \AA   & 8239.6 \AA   & 2 $\times 10^{-18}$ &
$\sim 20$ & 1.1189 \\

H$\gamma$ & 4341.69 \AA  & 9202.8 \AA   & 4 $\times 10^{-18}$ & $\sim 20$
& 1.1196 \\

\hline
\end{tabular}

\raggedright $^a$ Both restframe and observed emission line wavelength
are given in vacuum. The r.m.s.~statistical uncertainty in the line
centroiding is typically 0.3 \AA\ and the overall wavelength
calibration has an error of 0.1 \AA\ (1 $\sigma$). Statistical errors
on the redshift derived from an individual line is typically $\pm
0.0001$.

$^b$ Observed equivalent width, uncorrected for the continuum
contribution from the optical transient at the time of observation.

\end{table*}
\end{center}


\newpage


\begin{thebibliography}{35}
\expandafter\ifx\csname natexlab\endcsname\relax\def\natexlab#1{#1}\fi

\bibitem[{{Adelberger} \& {Steidel}(2000)}]{as00}
{Adelberger}, K.~L. and {Steidel}, C.~C. 2000, ApJ, 544, 218

\bibitem[{{Berger} {et~al.}(2002){Berger}, {Cowie}, {Kulkarni}, {Frail},
  {Aussel}, \& {Barger}}]{eck+02}
{Berger}, E., {Cowie}, L.~L., {Kulkarni}, S.~R., {Frail}, D.~A., {Aussel}, H.,
  and {Barger}, A.~J. 2002, submitted to ApJ

\bibitem[{{Berger} {et~al.}(2001){Berger}, {Diercks}, {Frail}, {Kulkarni},
  {Bloom}, {Sari}, {Halpern}, {Mirabal}, {Taylor}, {Hurley}, {Pooley},
  {Becker}, {Wagner}, {Terndrup}, {Statler}, {Wik}, {Mazets}, \&
  {Cline}}]{bdf+01}
{Berger}, E. {\it et al.} 2001, ApJ, 556, 556

\bibitem[{{Bloom} {et~al.}(2001){Bloom}, {Djorgovski}, \& {Kulkarni}}]{bdk01}
{Bloom}, J.~S., {Djorgovski}, S.~G., and {Kulkarni}, S.~R. 2001, ApJ, 554, 678

\bibitem[{{Bloom} {et~al.}(2002){Bloom}, {Kulkarni}, \& {Djorgovski}}]{bkd02}
{Bloom}, J.~S., {Kulkarni}, S.~R., and {Djorgovski}, S.~G. 2002, AJ, 123, 1111

\bibitem[{Bloom {et~al.}(2000{\natexlab{a}})}]{bdg+00}
Bloom, J.~S. {\it et al.} 2000{\natexlab{a}}, {{GCN}} notice 689

\bibitem[{Bloom {et~al.}(2000{\natexlab{b}})}]{bdd+00}
---. 2000{\natexlab{b}}, {GCN} notice 661

\bibitem[{{Bromm} \& {Loeb}(2002)}]{bl02}
{Bromm}, V. and {Loeb}, A. 2002, ApJ, 575, 111

\bibitem[{{Cardelli} {et~al.}(1989){Cardelli}, {Clayton}, \& {Mathis}}]{ccm89}
{Cardelli}, J.~A., {Clayton}, G.~C., and {Mathis}, J.~S. 1989, ApJ, 345, 245

\bibitem[{Castro {et~al.}(2000{\natexlab{a}})Castro, Diercks, Djorgovski,
  Kulkarni, Galama, Bloom, Harrison, \& Frail}]{cdd+00}
Castro, S.~M., Diercks, A., Djorgovski, S.~G., Kulkarni, S.~R., Galama, T.~J.,
  Bloom, J.~S., Harrison, F.~A., and Frail, D.~A. 2000{\natexlab{a}}, {GCN}
  notice 605

\bibitem[{Castro {et~al.}(2000{\natexlab{b}})}]{cdk+00}
Castro, S.~M. {\it et al.} 2000{\natexlab{b}}, {GCN} notice 851

\bibitem[{{Chary} {et~al.}(2002){Chary}, {Becklin}, \& {Armus}}]{cba02}
{Chary}, R., {Becklin}, E.~E., and {Armus}, L. 2002, ApJ, 566, 229

\bibitem[{de~Bernardis {et~al.}(2000)}]{dab+00}
de~Bernardis, P. {\it et al.} 2000, Nature, 404, 955

\bibitem[{{Djorgovski} {et~al.}(2001{\natexlab{a}}){Djorgovski}, {Frail},
  {Kulkarni}, {Bloom}, {Odewahn}, \& {Diercks}}]{dfk+01}
{Djorgovski}, S.~G., {Frail}, D.~A., {Kulkarni}, S.~R., {Bloom}, J.~S.,
  {Odewahn}, S.~C., and {Diercks}, A. 2001{\natexlab{a}}, ApJ, 562, 654

\bibitem[{{Djorgovski} {et~al.}(2001{\natexlab{b}}){Djorgovski}, {Kulkarni},
  {Bloom}, {Frail}, {Harrison}, {Galama}, {Reichart}, {Castro}, {Fox}, {Sari},
  {Berger}, {Price}, {Yost}, {Goodrich}, \& {Chaffee}}]{dkb+01}
{Djorgovski}, S.~G. {\it et al.} 2001{\natexlab{b}}, in Gamma-Ray Bursts in the
  Afterglow Era, Proceedings of the International workshop held in Rome, CNR
  headquarters, 17--20 October, 2000, ed. E.~Costa, F.~Frontera, \& J.~Hjorth
  (Berlin Heidelberg: Springer), 218

\bibitem[{{Fomalont} {et~al.}(2002){Fomalont}, {Kellermann}, {Partridge},
  {Windhorst}, \& {Richards}}]{fkp+02}
{Fomalont}, E.~B., {Kellermann}, K.~I., {Partridge}, R.~B., {Windhorst}, R.~A.,
  and {Richards}, E.~A. 2002, AJ, 123, 2402

\bibitem[{{Fukugita} {et~al.}(1995){Fukugita}, {Shimasaku}, \&
  {Ichikawa}}]{fsi95}
{Fukugita}, M., {Shimasaku}, K., and {Ichikawa}, T. 1995, PASP, 107, 945

\bibitem[{{Hogg} {et~al.}(1998){Hogg}, {Cohen}, {Blandford}, \&
  {Pahre}}]{hcbp98}
{Hogg}, D.~W., {Cohen}, J.~G., {Blandford}, R., and {Pahre}, M.~A. 1998, ApJ,
  504, 622

\bibitem[{Holland {et~al.}(2000)Holland, Thomsen, Bjornsson,
  {et~al.}}]{htab+00}
Holland, S., Thomsen, B., Bjornsson, G., {\it et al.} 2000, {{GCN}} notice 731

\bibitem[{{Kennicutt}(1992)}]{ken92}
{Kennicutt}, Robert~C., J. 1992, ApJ, 382, 310

\bibitem[{{Lamb} \& {Reichart}(2000)}]{lr00}
{Lamb}, D.~Q. and {Reichart}, D.~E. 2000, ApJ, 536, 1

\bibitem[{{Ledlow} {et~al.}(2002){Ledlow}, {Smail}, {Owen}, {Keel}, {Ivison},
  \& {Morrison}}]{lso+02}
{Ledlow}, M.~J., {Smail}, I., {Owen}, F.~N., {Keel}, W.~C., {Ivison}, R.~J.,
  and {Morrison}, G.~E. 2002, ApJ ({\it Letters}), 577, L79

\bibitem[{Metzger {et~al.}(2000)Metzger, Fruchter, Masetti, {et~al.}}]{mfm+00}
Metzger, M., Fruchter, A., Masetti, N., {\it et al.} 2000, {GCN} notice 733

\bibitem[{{Mirabal} {et~al.}(2002){Mirabal}, {Halpern}, {Kulkarni}, {Castro},
  {Bloom}, {Djorgovski}, {Galama}, {Harrison}, {Frail}, {Price}, {Reichart},
  {Ebeling}, {Bunker}, {Dawson}, {Dey}, {Spinrad}, \& {Stern}}]{mhk+02}
{Mirabal}, N. {\it et al.} 2002, ApJ, 578, 818

\bibitem[{{Oke}(1990)}]{oke90}
{Oke}, J.~B. 1990, AJ, 99, 1621

\bibitem[{{Osterbrock}(1974)}]{ost74}
{Osterbrock}, D.~E. 1974, {Astrophysics of gaseous nebulae} (San Francisco:
  W.~H.~Freeman and Co.)

\bibitem[{{Piro} {et~al.}(2002){Piro}, {Frail}, {Gorosabel}, {Garmire},
  {Soffitta}, {Amati}, {Andersen}, {Antonelli}, {Berger}, {Frontera}, {Fynbo},
  {Gandolfi}, {Garcia}, {Hjorth}, {Zand}, {Jensen}, {Masetti}, {M{\o}ller},
  {Pedersen}, {Pian}, \& {Wieringa}}]{pfg+02}
{Piro}, L. {\it et al.} 2002, ApJ, 577, 680

\bibitem[{{Rola} {et~al.}(1997){Rola}, {Terlevich}, \& {Terlevich}}]{rtt97}
{Rola}, C.~S., {Terlevich}, E., and {Terlevich}, R.~J. 1997, MNRAS, 289, 419

\bibitem[{Savaglio {et~al.}(2002)Savaglio, Fiore, Israel, Antonelli, Fontana,
  Stella, Paola, Stratta, Covino, Chincarini, Ghisellini, Saracco, Zerbi,
  Lazzati, Perna, Vietri, Frontera, Mereghetti, Meurs, \& Kawai}]{sfi+02}
Savaglio, S. {\it et al.} 2002, {GCN} notice 1633

\bibitem[{Schlegel {et~al.}(1998)Schlegel, Finkbeiner, \& Davis}]{sfd98}
Schlegel, D.~J., Finkbeiner, D.~P., and Davis, M. 1998, ApJ, 500, 525

\bibitem[{{Sheinis} {et~al.}(2002){Sheinis}, {Bolte}, {Epps}, {Kibrick},
  {Miller}, {Radovan}, {Bigelow}, \& {Sutin}}]{sbe+02}
{Sheinis}, A.~I., {Bolte}, M., {Epps}, H.~W., {Kibrick}, R.~I., {Miller},
  J.~S., {Radovan}, M.~V., {Bigelow}, B.~C., and {Sutin}, B.~M. 2002, PASP,
  114, 851

\bibitem[{{Stern} {et~al.}(2000){Stern}, {Bunker}, {Spinrad}, \&
  {Dey}}]{sbs+00}
{Stern}, D., {Bunker}, A., {Spinrad}, H., and {Dey}, A. 2000, ApJ, 537, 73

\bibitem[{{Taylor} {et~al.}(2000){Taylor}, {Bloom}, {Frail}, {Kulkarni},
  {Djorgovski}, \& {Jacoby}}]{tbf+00}
{Taylor}, G.~B., {Bloom}, J.~S., {Frail}, D.~A., {Kulkarni}, S.~R.,
  {Djorgovski}, S.~G., and {Jacoby}, B.~A. 2000, ApJ ({\it Letters}), 537, L17

\bibitem[{{Yost} {et~al.}(2002){Yost}, {Frail}, {Harrison}, {Sari}, {Reichart},
  {Bloom}, {Kulkarni}, {Moriarty-Schieven}, {Djorgovski}, {Price}, {Goodrich},
  {Larkin}, {Walter}, {Shepherd}, {Fox}, {Taylor}, {Berger}, \&
  {Galama}}]{yfh+02}
{Yost}, S.~A. {\it et al.} 2002, ApJ, 577, 155

\bibitem[{{Yun} \& {Carilli}(2002)}]{yc02}
{Yun}, M.~S. and {Carilli}, C.~L. 2002, ApJ, 568, 88

\end{thebibliography}
\end{document}